# How to Extract Energy Directly from a Gravitational Field

Fran De Aquino
Physics Department, Maranhao State University, S.Luis/MA, Brazil.

Gravity is related to *gravitational mass* of the bodies. According to the *weak* form of Einstein's General Relativity *equivalence principle*, the gravitational and inertial masses are equivalent. However recent calculations (gr-qc/9910036) have revealed that they are correlated by an adimensional factor, which depends on the incident radiation upon the particle. It was shown that there is a direct correlation between the radiation absorbed by the particle and its gravitational mass, independently of the inertial mass. This finding has fundamental consequences to Unified Field Theory and Quantum Cosmology. It was also shown that only in the absence of electromagnetic radiation this factor becomes equal to *one* and that, in specific electromagnetic conditions, it can be reduced, nullified or made negative. This means that there is the possibility of control of the *gravitational mass* by means of the incident radiation. This unexpected theoretical result was recently confirmed by an experiment (gr-qc/0005107). Consequently there is a strong evidence that the gravitational forces can be reduced, nullified and *inverted* by means of electromagnetic radiation. This means that, in practice we can produce *gravitational binari*es, and in this way to extract energy from a gravitational field. Here we describe a process by which energy can be extracted directly from any site of a gravitational field.

## INTRODUCTION

In a previous paper[1] we have shown that the general expression of correlation between gravitational mass $m_g$ and inertial mass $m_i$, is given by

$$m_g = m_i - 2\left\{\sqrt{1+\left[\frac{U}{m_i c^2}\sqrt{\frac{\varepsilon_r \mu_r}{2}\left(\sqrt{1+(\sigma/\omega\varepsilon)^2}+1\right)}\right]^2} - 1\right\}m_i \quad (1)$$

where $U$ is the electromagnetic energy absorbed by the particle; $\varepsilon$, $\mu$ and $\sigma$, are the electromagnetic characteristics of the outside medium around the particle in which the incident radiation is propagating. For an *atom* inside a body, the incident radiation on this atom will be propagating inside the body, and consequently, $\sigma = \sigma_{body}$, $\varepsilon = \varepsilon_{body}$, $\mu = \mu_{body}$. So, if $\omega \ll \sigma_{body}/\varepsilon_{body}$, equation above reduces to

$$m_g = m_a - 2\left\{\sqrt{1+\left[\frac{U}{m_a c^2}\sqrt{\frac{c^2 \mu_{body}\sigma_{body}}{4\pi f}}\right]^2} - 1\right\}m_a \quad (2)$$

where $m_a$ is the *inertial* mass of the atom.

This equation shows clearly that, *atoms* (or *molecules*) can have their *gravitational masses* strongly reduced by means of extra-low frequency (ELF) radiation.

We built a system to verify the effects of the ELF radiation on the gravitational mass of a body. The experimental setup and the obtained results were presented in a recent paper[2]. That experiment confirmed that the general expression of correlation between gravitational mass and inertial mass (Eq.1) is true. In practice, this means that the gravitational forces can be reduced, nullified and *inverted* by means of electromagnetic radiation. Therefore we can build *gravitational binari*es, and in this way to extract energy from any site of a gravitational field.

In this work we present a system to extract energy directly from a gravitational field, which is basically a gravitational binary produced by means of gravity control. We named this system : *Gravitational Motor*.

# THE GRAVITATIONAL MOTOR

The experimental system is shown in Figure 1. It is based on the system-G presented in reference[2].

The *annealed pure iron* utilized in the system-G, is now in *three* tubes in the rotor of the motor (2 externals and 1 internal, see fig.1a). It has an electric conductivity $\sigma_i=1.03\times10^7$S/m, magnetic permeability[3] $\mu_i=25000\mu_0$, thickness 20 mm ( to absorb totally the ELF radiation produced by the antenna). Here the ELF antenna is encapsulated by a *ceramic ferromagnetic material*, which must has $\sigma_p \approx 10$ S/m; $\mu_p \approx$ **7500$\mu_0$**.. Note that, the relative permeability is 100 times greather than the relative permeability of the *iron powder* of the System-G. The antenna physical length is $z_0 = 12$ m, (see Fig.1c). The power radiated by the antenna can be calculated by the well-known *general* expression, for $z_0 \ll \lambda$ :

$$P = (I_0 \omega z_0)^2 / 3\pi\varepsilon v^3 \{[1+ (\sigma/\omega\varepsilon)^2]^{½} +1\} \quad (3)$$

where $I_0$ is the antenna current amplitude ; $\omega = 2\pi f$ ; $f =0.6$Hz ( Note that here the frequency is 100 times less than in the system-G) ; $\varepsilon =\varepsilon_p$ ; $\sigma =\sigma_p$ and $v$ is the wave phase velocity in the *ceramic ferromagnetic material* ( given by Equation1.02 , in reference [1] ). The radiation efficiency $e = P / P+P_{ohmic}$ is nearly 100%.

Each atom of the **annealed pure iron tubes** absorbs an ELF energy $U=\eta P_a/f$, where $\eta$ is a particle-dependent absorption coefficient (the maxima $\eta$ values occurs, as we know, for the frequencies of the atom's *absorption spectrum* ) and $P_a$ is the incident radiation power on the atom ; $P_a=DS_a$ where $S_a$ is the atom's *geometric* cross section and $D=P/S$ the radiation power density on the iron atom ( $P$ is the power radiated by the antenna and $S$ is approximately equal to the lateral area of the encapsulated antenna. We are assuming that $D$ is approximately constant inside the iron tubes). Thus, we can write :

$$U =\eta S_a(I_0 z_0)^2 \omega/3S\varepsilon_i v^3\{[1+(\sigma_i/\omega\varepsilon_i )^2]^{½}+1\} \quad (4)$$

Consequently, according to Eq.(1) , for $\omega\ll\sigma_i/\varepsilon_i$ , the gravitational masses of these iron *atoms*, under these conditions, will be given by :

$$m_g= m_a-2\{[1+6.2\times10^{-4}I_0^4 S^{-2} ]^{½}-1\}m_a \quad (5)$$

Equation above shows that the *gravitational masses* ($m_g$) of the atoms of the annealed pure iron can be *nullified* for $I_0 \cong 6.7$A if $S=1$m$^2$. Above this critical value the gravitational masses becomes negatives .

The iron tubes of the rotor start to spin when the symmetry of the gravitational forces acting on them is broken .i.e., when the gravitational forces acting on the part submitted to the ELF radiation start to be reduced ( the left side of the iron tubes, see Fig.1d) .

Let us assume that $m_i$ is the *inertial mass* of each *half* of a iron tube and $m_g$ the *gravitational mass* of the half submitted to the ELF radiation, and that $m_g$ was made negative, in such way that $m_g=-Nm_i$, $N>0$. Then we can consider that on the left side of the iron tube the *weight* is <u>inverted</u> and its intensity is $|m_g|g=Nm_ig$, in agreement with Eq.(2.05) of the reference [1]. On the other hand, on the right of the tube the weight it is preserved and equal to $m_ig$ .Therefore, it is easy to see that the *tangential* acceleration $a_T$ of the tube will be equal to $g$. Thus, the *medium angular acceleration* $\alpha$ of the tube will be given by:

$$\alpha = \frac{a_T}{r} = g/r \quad (6)$$

where $r=(r_1+r_2)/2$; $r_1$ and $r_2$ are respectively, the external and internal radius of the tube.

Consequently, the *Torque* T, will be given by: $T = I\alpha = Ig/r$, where $I = \frac{1}{2}|M_g|(r_1^2 + r_2^2)$ is the moment of inertia of the tube.

Note that the equation of the moment of inertia contains $|M_g|$ instead of the inertial mass. This is a consequence of the new expression for the kinetics energy,

$$K = \frac{1}{2}|M_g|v^2, \qquad (7)$$

which is obtained from the equation of the Total Energy (Eq.(2.07) of the reference [1]). It is known that we can write $K = \frac{1}{2}(\Sigma m_j r_j^2)\omega^2$ where $(\Sigma m_j r_j^2) = I$ is called moment of inertia of the body in respect to its rotation axis. Consequently, due to the equation(7), the $m_j$ in the equation above, refers now to the *gravitational masses*.

Finally, we can write $K=\frac{1}{2}I\omega^2=$ (Force)•(Displacement) = $(|m_g|g+m_ig)$•$(2\pi r)$. Consequently, we obtain

$$\omega = \sqrt{8\pi g \left(\frac{r_1 + r_2}{r_1^2 + r_2^2}\right)} \qquad (8)$$

Thus, the *power* of the spinning iron tube gives

$$P = T\omega = |M_g|\sqrt{2\pi g^3 \left(\frac{r_1^2 + r_1^2}{r_1 + r_2}\right)} \qquad (9)$$

where $|M_g| = |m_g| + m_i = (N+1)m_i$.

Therefore, if we assume that the *internal* iron tube of the rotor has the followings characteristics: $r_1=0.10$m; $r_2=0.08$m;L=H=0.50m( L is the length of the tube); $\rho=7.8\times10^3$kg/m$^3$ (pure iron), and that $S=0.4$m$^2$ and $I_0=29$A, then equation(5) tells us that $N\cong10^2$. Thus, according to equation(9) we obtain

$$P = 57.4 \text{ Kw} \cong 77 \text{ HP}$$

Note that this power refers solely to the power of the *internal* iron tube. The rotor presented in Fig.1, still has *two external* iron tubes with $r_1=0.15$m; $r_2=0.13$m;L=0.24m; that provide more 51.4Kw each one. Therefore, the total power of the motor gives

$$P=57.4+2\times51.4=160.2\text{Kw} \cong \mathbf{214 \text{ HP}}$$

It is easily seen that the only difficulty to build the motor, it is to get the **ceramic ferromagnetic material** with the previously mentioned characteristics. i.e., $\sigma_p \approx 10$ S/m ; $\mu_p \approx \mathbf{7500}\mu_0$.

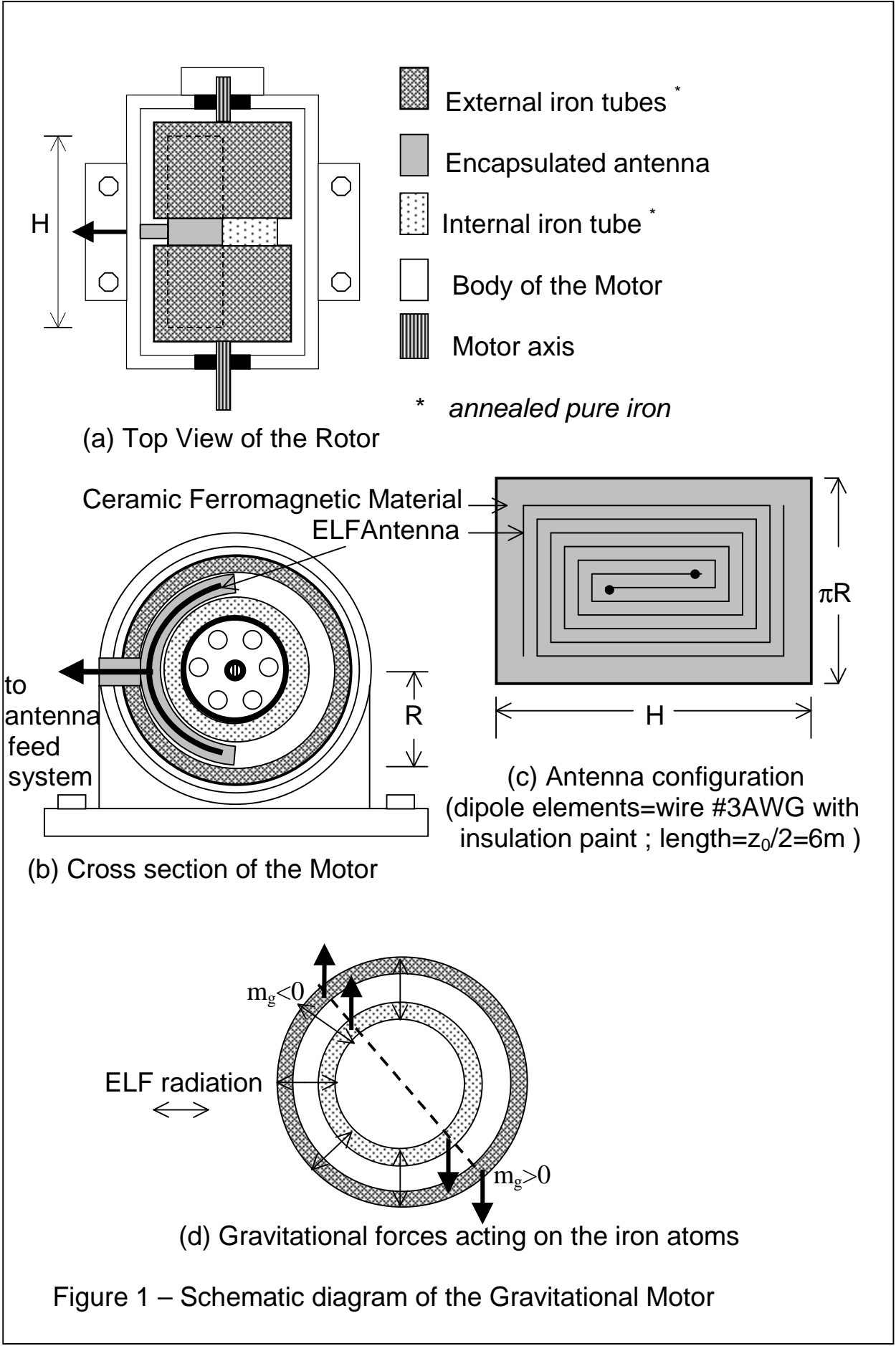

Figure 1 – Schematic diagram of the Gravitational Motor